# THE STRIKING NEAR-INFRARED MORPHOLOGY OF THE INNER REGION IN M100

J.H. Knapen[1,2], J.E. Beckman[2], I. Shlosman[3,4], R.F. Peletier[2,5,6], C.H. Heller[3,7], and R.S. de Jong[5]

[1] Département de Physique, Université de Montréal, C.P. 6128, Succursale Centre Ville, Montréal (Québec), H3C 3J7 Canada (present address).
[2] Instituto de Astrofísica de Canarias, E-38200 La Laguna, Tenerife, Spain
[3] Department of Physics and Astronomy, University of Kentucky, Lexington, KY 40506-0055, USA
[4] Gauss Foundation Fellow
[5] Kapteyn Astronomical Institute, Postbus 800, NL-9700 AV Groningen, the Netherlands.
[6] Royal Greenwich Observatory, Apartado 321, E-38700 Santa Cruz de La Palma, Spain.
[7] Universitäts Sternwarte Göttingen, Geismarlandstraße 11, D-37083 Göttingen, Germany (present address).

## Abstract

New optical and near-infrared (NIR) $K$-band images of the inner 3 kpc region of the nearby Virgo spiral M100 (NGC 4321) display remarkable morphological changes with wavelength. While in the optical the light is dominated by a circumnuclear zone of enhanced star formation, the morphological features in the $2.2\mu$m image correspond to a newly discovered kpc-size stellar bar, and a pair of *leading* arms situated inside an ovally-shaped region. Analysis of the $K$ image confirms its symmetry: only a very small percentage of the flux, some 5%, is emitted in antisymmetric structures. This indicates that the overall morphology observed in the NIR is dominated by a global density wave. Making a first-order correction for the presence of localized dust extinction in the $K$ light using the $I - K$ image, we find that the observed leading arm morphology is not caused or enhanced, but in fact slightly hidden by dust. Possible mechanisms responsible for the optical and NIR morphology are discussed, and tests are proposed to discriminate between them. Our dynamical conclusions are supported with an evolutionary stellar population model reproducing the observed optical and NIR colors in a number of star forming zones. We argue that the observed morphology is compatible with the presence of a pair of inner Lindblad resonances in the region, and show this explicitly in an accompanying paper by detailed numerical modeling. The phenomena observed in NGC 4321 may provide insight into physical processes leading to central activity in galaxies.

*Subject headings:* hydrodynamics — galaxies: individual (M100, NGC 4321) — galaxies: evolution, ISM, starburst, structure

# 1 Introduction

The possibility of multiple inner Lindblad resonances (ILRs) in the central regions of bar-driven spiral galaxies has been predicted by Sanders & Huntley (1976) and Lynden-Bell (1979). This conjecture has been supported so far by circumstantial evidence only because of the uncertainties involved in determining the pattern speed of the bar (*e.g.* Arsenault *et al.* 1988; Kenney *et al.* 1992; Kenney, Carlstrom & Young 1993). Besides the shape of rotation curves in the inner parts of galaxies, the main arguments in favor of ILRs include dust lanes along the leading edges of the bars (Prendergast 1962; Athanassoula 1992) and so-called nuclear rings of star formation (SF) (Simkin, Su & Schwarz 1980; Combes & Gerin 1985; Buta 1986; Athanassoula 1992; Buta & Crocker 1993; Shaw *et al.* 1993; Elmegreen 1994). Rings of molecular gas are found in such regions as well (*e.g.* Sofue 1991).

NGC 4321 (M100) is a late-type barred galaxy with a prominent nuclear ring of SF claimed to lie in the vicinity of the ILR(s) (Pierce 1986; Arsenault *et al.* 1988). It is favorably inclined at $30°$ ($\pm 3°$) and has well-defined spiral arms. The $60''$ semimajor axis stellar bar is visible only in the NIR (Pierce 1986). In this *Letter* we use a high-resolution $K$-band image of NGC 4321 to analyze the NIR morphology in the central $20''$ radius region ($12''$=1 kpc at the assumed distance of $D = 17.1$ Mpc; Freedman *et al.* 1994) and to compare it with an $I - K$ dust extinction map and an H$\alpha$ image. In Paper II (Knapen *et al.* 1995) we strengthen our conclusions by demonstrating numerically the close relationship between the observed morphology and the underlying dynamics of the region: the main zones of SF all appear to lie deep inside the resonance zone and close to the inner ILR (IILR).

# 2 Observations

A NIR $K$-band ($2.2\mu$m) image was obtained with the IRCAM3 camera on UKIRT. The projected pixel size of $0\rlap.{''}286$ was adequate to sample the seeing of $0\rlap.{''}8$ FWHM. The field of view was larger than the $40'' \times 40''$ shown in Fig. 1a. A broad-band $I$ ($0.8\mu$m) and an H$\alpha$ continuum subtracted image of the center of NGC 4321 were obtained with the 4.2m WHT, using the auxiliary port and TAURUS cameras. The resolution is around $0\rlap.{''}8$, at pixel sizes of $0\rlap.{''}1$ ($I$) and $0\rlap.{''}27$ (H$\alpha$). Fig. 1c shows our H$\alpha$ image (same scale as Fig. 1a), and Fig. 1d shows a $V$-band HST image (after refurbishment) obtained from the HST archive. An $I - K$ color index image (Fig. 2) was made by dividing the $K$ by the $I$ image, after alignment; this is chiefly a dust extinction map (see below).

# 3 Understanding the NIR Morphology

In our $K$-band image (Fig. 1a), the notable features are i) A round small bulge of $4''$ (330 pc) effective diameter; ii) An inner bar with major axis of $9''$ (750 pc) along position angle (PA) $111° \pm 1°$ (anti-clockwise from N) and ellipticity $\sim 0.6$. This PA is equal within the fitting errors to that of the 5 kpc (radius) stellar bar (Pierce 1986; Knapen *et al.* 1993). iii) An oval ring-like region between $\sim 10'' - 22''$ whose minimal ellipticity ($0.13 \pm 0.02$ at $18''$) and PA=$153° \pm 3°$ are the same as those of the outer disk: it most probably lies in the disk plane. The $K$-isophotes (and $I$-isophotes, Paper II) are skewed progressively towards the PA of the bar at smaller and larger radii from the "ring". iv) Two strong maxima (K1 and K2, Fig. 1a) of $K$ emission lying on a straight line through the nucleus and equidistant from it, at $\sim 600$ pc. The PA of this line is $114° \pm 1°$, *i.e.* slightly skewed with respect to the bar *in* the direction of rotation. v) Two *leading* arms connecting the bar to the K1/2 peaks, not seen in the optical or even the HST image (Fig. 1d).

In the visible (Figs. 1c,d), the most striking features are the tightly wound, segmented armlets. The $I - K$ (Fig. 2) and the HST $V$-band (Fig. 1d) images show two strong opposing dust lanes swinging inwards from outside the mapped zone. These and the accompanying spiral arms can be traced from the main disk to $\sim 600$ pc from the center; the arms intensify dramatically on entering the central kpc. Peak K1 appears to be an active SF zone coinciding with strong H$\alpha$ and $V$ emission. In $I - K$ it is seen surrounded by a complete dust shell. Peak K2 does not coincide perfectly but is sandwiched by its H$\alpha$ counterpart and embedded in dust. Peaks H$\alpha$3/4, prominent in H$\alpha$, are fading in $I$ (Paper II) and largely absent in $K$.

The $I - K$ color index map was used to estimate the corrections in the $K$ image due to dust extinction, assuming that deviations from an average $I - K$ color (Fig. 2) are due solely to dust and that the Galactic extinction law applies (Knapen *et al.* 1991; Jansen *et al.* 1994). The precise use of this law in any given case requires a knowledge of the geometry of the mixture of dust and stars, but because the $K$ extinction due to dust is small here (see below) any errors entailed in this assumption are of second order. Although, in general, an optical-NIR color index image shows both dust extinction and stellar population effects, by choosing $I - K$ we have minimized the population component (*e.g.* Rix & Rieke 1993), while retaining significant extinction ($A_I/A_V = 0.48, A_K/A_V = 0.12$).

We start with subtracting an average color from the $I - K$ map. We ignore two types of systematic errors: a contribution due to smoothly distributed dust (not altering the morphology), and effects due to population changes (which will be small in any case). Using the Galactic values of $A_K/A_V$ and $A_I/A_V$, we corrected the $K$ image by subtracting 25% from the resulting $I - K$. This adds only $\sim 3\%$ to the flux in $K$, and confirms that very little emission is lost at 2.2$\mu$m due to localized patches of dust. The dust-corrected $K$ image is shown in Fig. 1b at the same contour and color levels as the uncorrected $K$ image (Fig. 1a). Differences between Figs. 1a,b do reveal local effects of dust in the $K$ contours, notably in the leading arms which become more pronounced; hence they are not artifacts caused by dust.

To measure the degree of symmetry in the $K$ image, we have decomposed it into symmetric ($S$) and antisymmetric ($A$) parts (Elmegreen, Elmegreen & Montenegro 1992; Knapen 1992). An $S$ image (Fig. 3a) contains the two-fold symmetric flux in the original image, and $A$ (Fig. 3b) contains the flux that is antisymmetric (deprojection is not necessary, since this is also a two-fold symmetric operation). Almost all $K$ emission arises in parts of the central zone that are symmetrically placed with respect to the nucleus. Within a square box of $30''$ on a side centered on the nucleus, roughly the area of the inner structure, only some 5% of the $K$ flux is in $A$, whereas the vast majority of flux, 95%, is emitted from $S$ structures. This is direct proof that the underlying mechanism responsible for the dominant morphology must be global (see Block *et al.* 1994 for similar discussion). Comparison of the $K$ image with the bluer images shows that local perturbations produce regions of SF which are not always perfectly symmetrically placed in the inner region. For instance, in H$\alpha$ only some 66% of the flux in the inner 30$''$ is emitted from $S$ structures. Most of the antisymmetric flux in H$\alpha$ is emitted by strong H II regions stochastically placed in the inner spiral, although extinction by dust may slightly influence our findings in H$\alpha$.

The main contribution to the NIR luminosity of a quiescent galaxy comes from an old stellar population represented by K and M giants which follow the overall mass distribution (*e.g.* Frogel 1985). A bar which stands out in $K$ is hence dominated by these stars, older than $10^9$ yrs and populating orbits aligned with the bar. Only if a *double* ILR is present within the bar, stellar orbits between the ILRs are allowed to be oriented along the minor axis of the bar (Sanders & Huntley 1976), an effect easily understood within the framework of forced oscillations (*e.g.* Landau & Lifshitz 1969). We note that in the presence of even a moderate strength bar the linear analysis based on the rotation curve is insufficient to test for the presence of ILRs (*e.g.* Heller & Shlosman 1994).

At least three of our observational results speak in favor of a double ILR in NGC 4321. First, the $K$ isophotes, starting at around 30$''$ from the center, show gradual skewing towards the PA of the ovally-shaped ring at 18$''$ and then back towards the bar at smaller radii, as partly shown in Fig. 1 (also Paper II, Fig. 8). At $\sim 10''$ they are again aligned with the bar, within the margin of error. The only complication (addressed below) is that such a twisting is anticipated in the gas flow rather than in stars: stellar orbits change their PA abruptly at each resonance, gas orbits can do so only gradually (Sanders & Huntley 1976; Huntley, Sanders & Roberts 1978). Second, a pair of dust lanes, long believed to delineate the large scale shocks in the gas flow of barred galaxies (Prendergast 1962), exists in the 5 kpc bar (partly shown in Fig. 2). Their large offset with respect to the bar's major axis is indicative of a double ILR in this case. Third, a pair of leading armlets is seen clearly in $K$ and is not an artifact of dust obscuration. Both the leading and the "conventional" trailing arms start in K1 and K2 which stand out in $K$ and are remarkably symmetric with respect to the nucleus (Figs. 1, 3). Finding two systems of spiral arms, leading and trailing, interacting close to an ovally-shaped structure suggests resonance effects in the neighborhood of the IILR (*e.g.* Sanders & Huntley 1976; Schwarz 1984; Paper II). In fact, the K1/2 peaks around 600 pc from the center lie just outside the region of a steep rotational velocity gradient (Arsenault *et al.* 1988). This radius delineates also the inner boundary of SF activity in the circumnuclear "ring" as given by H$\alpha$ intensity. Of the four main SF sites seen in H$\alpha$ (Fig. 1c), H$\alpha$1 and H$\alpha$2 have clear counterparts in $K$ (K1/2 peaks), whereas H$\alpha$3 and H$\alpha$4 are hardly seen in $K$. What is then the relationship between the gas flow to the $K$ light in the circumnuclear region of NGC 4321?

There are two possible ways to relate the twisting of $K$ isophotes within the stellar bar to the gas dynamics in the presence of ILRs. First, some of the old stellar population in the bar can be affected by gas self-gravity and populate gas orbits (Shaw *et al.* 1993). Alternatively, there is a non-negligible contribution from young massive stars, either OB stars or K and M supergiants, both being injected along the gas orbits. The orbital time scale of a few$\times 10^7$ yrs characteristic of azimuthal mixing then nicely distinguishes an older population of $> 10^9$ yrs from a younger one of $< 10^7$ yrs, on kinematical grounds.

What is the origin of the K1/2 peaks in the NIR? Their symmetry with respect to the bar and the nucleus, as well as their clearly being the locus points for leading and trailing arms, proves that K1/2 are part of the global density wave driven by the bar. On the other hand, peak K1 coincides with peak H$\alpha$1 which is a SF region. Although K2 does not correspond directly to an H$\alpha$ peak, it is surrounded by H$\alpha$ emission and is probably a younger starburst still shrouded in dust. In an attempt to quantify this, and to determine the stellar age, we have measured representative colors $U - V$ and $V - K$ in small apertures centered on the SF regions K1/2, H$\alpha$3/4 (Fig. 1c), and on the nucleus. We plot the colors in Fig. 4 and show vectors indicating (Galactic) reddening, and scattering plus absorption (following the "dusty galaxy" model from Witt, Thronson & Capuano 1992). The four SF regions are close together in this $U - V$ vs. $V - K$ diagram, revealing similar colors (errors are $\sim 0.1$ mag).

We compare the observed colors with photometric models. First we show a model of a $10^7$ year-old coeval starburst (Charlot & Bruzual 1991). An age of $\log(t) = 7.2 - 7.3$ (15-20 Myr), when $V - K$ is large and $U - V$ still negative, indicates the epoch of red supergiants. Two composite models were constructed by adding a varying fraction of young population of $\log(t) = 6.65$ (full line) and $\log(t) = 7.6$ (dotted line) to the population of the center, which is assumed to be the underlying old population. The numbers beside the lines indicate the relative fraction of young-to-old light in $V$. For $\log(t) = 6.65$ the models are too blue in $U - V$ for a given $V - K$, but for $\log(t) = 7.6$ they lie close to the observed colors of the SF regions.

One infers two extreme solutions. In the first, all regions consist of young stars of $\sim$15 Myr, and the different colors of the regions are due to a combination of absorption/scattering by dust, with regions from H$\alpha$3 through to K2 being increasingly dusty. In the second, the regions show different colors due to a change of the fraction of young-to-old stars in a composite model (thus some stars are older than 40 Myr), with regions K2 to H$\alpha$3 containing an increasing fraction of young stars. It is not possible to decide which of the two extremes, or which combination of the two, is the correct one from the presented colors alone. Our analysis favors that all four regions K1/2 and H$\alpha$3/4 are indeed SF regions, where K1/2 are in the middle of dust lanes, while H$\alpha$3/4 locally define the spiral arms. We also noted that K1 may be in a more advanced state of its starburst activity than K2. Where K2 is still hidden in the dust, K1 has pushed it aside and has emerged as a strong emitter in blue light (Figs. 1d, 2).

These considerations lead us to believe that dust is the major cause of the observed colors. In this interpretation most of the stars in all the four SF regions are equally young ($\sim$15 Myr), with H$\alpha$3 slightly less and K2 considerably more dusty than H$\alpha$4 and K1. More observations are needed to test our conclusions, and to determine whether (and how much) the old stellar population affects the colors. Observational effort should be focused on obtaining additional ($J$, $H$) NIR imaging, or spectral information of selected regions.

Thermal emission by hot dust heated by radiation from O stars, known to be an important factor in starburst regions (*e.g.* Scoville *et al.* 1988; Telesco 1993), might help to explain why the leading arms show up so clearly in $K$ and not in any of the other tracers. One might expect that, as in the case of K1/2, at least some of the massive SF activity causing the dust to emit shows up in the blue or H$\alpha$. Dust is present in the region where the leading arms are found ($I - K$ map), and also some very fine structure in H$\alpha$ is seen there (*e.g.* a faint and very narrow bridge connecting K1 with the nucleus), but further NIR observations are needed to decide on the issue of hot dust.

## 4   Conclusions

The inner region of the barred spiral NGC 4321 shows remarkably different morphology in the optical and the NIR. Whereas in the optical it is dominated by two spiral arms lying in an ovally-shaped region of enhanced SF, a $K$-band image reveals an inner bar aligned with the 5 kpc stellar bar and a pair of leading arms emerging from its ends. Neither feature is observed directly in the optical. Correction for dust extinction shows that these leading arms are not caused or enhanced by dust. Two "hot-spots" (K1/2) are seen in $K$ situated remarkably symmetrically with respect to the nucleus and slightly ahead of the stellar bar. We identify these with the loci of trailing and leading shocks inducing SF there. Combination of the $K$ image with others at shorter $\lambda$ shows a morphology characteristic of a double ILR. This is explicable via a dynamical model (Paper II), whose predictions can be tested kinematically, *e.g.* with mm wave interferometry, or Fabry-Pérot imaging spectroscopy. Comparing the symmetry of the $K$ image with that of the tracers of young stars (notably H$\alpha$) we conclude that the underlying pattern of stars, as seen in $K$, is highly symmetric and dominated by a global density wave, with SF occurring in the "ring" in a more stochastic way.

Using simple stellar population models to study the colors of four main SF regions in the "ring", we conclude that the major cause for the color differences between them is varying amounts of dust. In particular, one of the two "hot-spots" (K2) seen in $K$ is very dusty. It is a starburst that has not yet broken through the dust, and thus at a less advanced stage in its evolution as its counterpart K1, which has pushed aside the dust and is surrounded by a complete dust shell. Additional insight into the bar-driven inflow of NGC 4321 is provided by our numerical modeling in Paper II.

NGC 4321 is a nuclear starburst induced and maintained by a global bar-driven density wave. The location of the starburst in the circumnuclear "ring" is related to the slowing down of the radial gas inflow in the presence of ILRs (Shlosman, Frank & Begelman 1989). Understanding the details of such radial flows in barred galaxies may well shed light on the origin and fueling of active galactic nuclei.

*Acknowledgements.* The WHT is operated on the island of La Palma by the RGO in the Spanish Observatorio del Roque de los Muchachos of the IAC. The UK IR Telescope is operated by the RO in Edinburh on behalf of the SERC. Partly based on observations made with the NASA/ESA HST, obtained from the data archives at the STScI (operated by the AURA under NASA contract NAS 5-26555). JHK and JEB acknowledge DGICYT grant PB91-0525. IS acknowledges support from the Gauss Foundation, the IAC (P3/86), NASA grant NAGW-3839, and the CCS at the University of Kentucky.

# Figure Captions

**Figure 1a (Color Plate 1).** 2.2$\mu$m $K$-band image of the inner $40''$ (3.3 kpc) of NGC 4321. Contour levels: 16.7, 16.3, 15.9, from 15.6 to 14.6 in steps of 0.2, 14.2, 13.6, and 13.0 $K$-mag arcsec$^{-2}$. The contour plot is overlaid on a false-color representation of the same image, where red and white shades represent brighter areas. The center position at RA(1950)= $12^h 20^m 22\overset{s}{.}9$, dec= $16° 06' 00''$, is indicated with a cross. The zones K1 and K2 are SF complexes located at the loci of trailing and leading spiral arms (see text). The dashed line marks the PA of the major axis of the 5 kpc stellar bar.

**Figure 1b (Color Plate 1).** $K$ image as in Fig. 1a, but corrected for dust extinction using an $I - K$ color index image and assuming a Galactic extinction law. Contours and colors as in Fig. 1a.

**Figure 1c (Color Plate 1).** Inner part of continuum subtracted H$\alpha$ image of the same region as Fig. 1a. Image ($0\overset{''}{.}8$ resolution) obtained with the WHT at La Palma. Contour levels 0.7, 1.5, 3.0, 5.9, 9.3, 14.8, 19.7, 24.6 and 29.5$\times 10^{36}$ erg s$^{-1}$. Color representation of the same H$\alpha$ image shows brighter regions as red and white. SF regions K1 and K2, and H$\alpha$3 and H$\alpha$4 are indicated.

**Figure 1d (Color Plate 1).** WFPC-II HST image in the $V$-band of the same region as in Figs. 1a, b and c. Brighter regions are indicated as red and white. Contours of the $K$ image as in Fig. 1a are overlaid.

**Figure 2 (Black/White Plate 2).** $I - K$ color index image at $\sim 0\overset{''}{.}8$ resolution of the same region shown in Fig. 1, obtained by combining the UKIRT image (Fig. 1a) with an $I$-band CCD image from the WHT. Grey scales correspond to a range from 1.4 mag (black) to 3.1 mag (white), so that "bluer" features (relatively more emission at shorter wavelength, *i.e.* in $I$) are lighter, and "redder" features (relatively more emission at longer wavelength, *i.e.* in $K$) are darker in the Figure. Dust absorption features, notably dust lanes, thus show up darker than their surroundings.

**Figure 3.** (**a**) Symmetric ($S$) and (**b**) antisymmetric ($A$) part of the $K$-image shown in Fig. 1a. Contours as in Fig. 1a, plus an additional (outer) contour corresponding to 17.1 $K$-mag arcsec$^{-2}$. The sum of the $S$ and the $A$ images yields exactly the original $K$-band image. PA of the large-scale bar and positions of $K$ "hot spots" K1/2 are indicated as in Fig. 1a.

**Figure 4.** $U - V$ vs. $V - K$ color-color diagram showing the colors of four prominent SF regions: K1, K2, H$\alpha$3 and H$\alpha$4 (asterisks), and the center (marked C). Error bars of 0.1 mag in each color are indicated. Arrows at the bottom left: "Galactic" reddening caused by $A_V = 0.5$ (dashed arrow) and effects of scattering+absorption from the "dusty galaxy" model (Witt *et al.* 1992), for $\tau_V = 0.5$ and 2.0 (full line arrow). Dashed-dotted line: model of a $10^7$ year-old coeval starburst, small numbers indicate log($age$). Full and dotted lines: composite models of old and young populations, $\log(t) = 6.65$ and 7.6 respectively, numbers indicate relative fraction of young to old light in $V$.